\documentstyle[prd,aps,psfig]{revtex}

\newcommand\be{\begin{equation}}
\newcommand\ee{\end{equation}}
\newcommand\bea{\begin{eqnarray}}
\newcommand\eea{\end{eqnarray}}
\newcommand\ba{\begin{array}}
\newcommand\ea{\end{array}}
\begin{document}
\draft
\input epsf
\def\la{\mathrel{\mathpalette\fun <}}
\def\ga{\mathrel{\mathpalette\fun >}}
\def\fun#1#2{\lower3.6pt\vbox{\baselineskip0pt\lineskip.9pt
        \ialign{$\mathsurround=0pt#1\hfill##\hfil$\crcr#2\crcr\sim\crcr}}}

\twocolumn[\hsize\textwidth\columnwidth\hsize\csname
@twocolumnfalse\endcsname

\title{Gauge Unification and the Supersymmetric Light Higgs Mass}
\author{Jose Ram\'on Espinosa$^{(1)}$ and Mariano Quir\'os$^{(2)}$}
\address{\phantom{ll}}
\address{$^{(1)}${\it CERN TH-Division, CH-1211 Geneva 23, Switzerland}}
\address{$^{(2)}${\it IFAE, UAB 08193 Bellaterra, Barcelona, Spain}}
\address{and}
\address{{\it IEM, CSIC Serrano 123, 28006 Madrid, Spain}}
\date{\today}
\maketitle
\begin{abstract}
We consider general supersymmetric models with: 
a) arbitrary matter content; and, b) gauge coupling Unification 
near the String scale $\sim 10^{17}$ GeV,
and derive the absolute upper limit on the mass of the lightest 
Higgs boson. For a top-quark mass 
$M_t$= 175 GeV, and depending on the supersymmetric parameter 
$\tan\beta$, this mass bound can be as high as $\sim$ 200 GeV.
\end{abstract}
\vskip2pc]

Low-energy supersymmetry \cite{susy} is a key ingredient of the 
candidate best qualified models to supersede the Standard Model (SM) at
energies beyond the TeV range. The extensive experimental search of the
(super)-partners of SM elementary particles predicted by supersymmetry
(SUSY) has been so far unsuccessful, challenging \cite{ft},
with the rise of experimental mass limits, the naturalness and relevance
of SUSY for the electroweak scale physics. 

In this context, the sector of the theory responsible for electroweak
symmetry breaking has a special status. While all superpartners of the
known SM particles can be made heavy simply by rising soft SUSY-breaking
mass parameters in the model, the Higgs sector necessarily contains a
physical Higgs scalar whose mass does not depend sensitively on the
details of soft masses but is fixed by the scale of electroweak symmetry
breaking. This important fact follows simply from the spontaneous
breaking of the electroweak gauge symmetry \cite{genb,coes} and it is not
specific to supersymmetric models. More precisely, the general
statement is that some Higgs boson must exist whose mass-squared 
satisfies $m_h^2\leq \lambda v^2$, where $v$ is the electroweak scale
($v=\ 174.1$ GeV) and $\lambda$ is the dimensionless quartic coupling of
some Higgs state in the model. In other words, the mass of the light
Higgs can be made heavy only at the expense of making the
coupling $\lambda$ very strong. 
The role of supersymmetry is to fix $\lambda$ in some
models. For example, in the Minimal Supersymmetric Standard Model (MSSM),
which includes two Higgs doublets
\be
H_1=\left(\begin{array}{c}
H_1^o\\
H_1^-
\end{array}\right),\hspace{1cm}
H_2=\left(\begin{array}{c}
H_2^+\\
H_2^o
\end{array}\right),
\ee
to give masses to quarks and leptons, $\lambda$ is related 
to the $SU(2)_L\times U(1)_Y$ gauge couplings ($g$ and $g'$
respectively) and the following (tree-level) bound on the mass of
the lightest Higgs boson holds
\be
m_h^2\leq M_Z^2\cos^22\beta,
\ee
where $\tan\beta=\langle H_2^0\rangle/\langle H_1^0\rangle$.
This represents a very stringent prediction which, as is well known,
gets significantly relaxed when radiative corrections to $\lambda$ are
included \cite{radcor1,radcor2,radcor3}. These corrections depend
logarithmically on the
soft-masses and push the upper mass limit for the lightest
Higgs boson of the MSSM up to $125$ GeV (for a top-quark 
mass $M_t=175$ GeV).

The fact that $\lambda$ is calculable in terms of gauge couplings in the
MSSM is due to supersymmetry and to the fact that the
superpotential does not contain cubic terms of the form 
$W=h_X XH_iH_j$
(with $i,j=1,2$) as no $X$ field exists with the appropriate quantum
numbers
to form a gauge invariant object. In extended models, the presence of
such fields and couplings modifies the quartic Higgs self-interactions
which can ultimately have an impact on the tree-level upper bound on
$m_h^2$, which receives corrections proportional to $h_X^2v^2$. 
The Yukawa
coupling $h_X$ is unknown but asymptotically non-free, 
and so it can be
bounded from above if it is further required to remain in the
perturbative regime up to some large energy scale (where GUT, String or 
Planck physics takes over).

As there is no reason to believe that low-energy supersymmetry is
realized in nature in the form of the MSSM, the experimentalist willing
to test SUSY via Higgs searches would like to know what is the absolute
general upper limit that should be reached on $m_h$ to rule 
out low-energy SUSY. This is the important question we set 
ourselves to answer in this letter.

{\bf 1.} To give a precise answer we have to assume that all couplings in
the theory remain perturbative up to a very large energy scale. This is
particularly well motivated in low-energy SUSY models from the successful
unification of gauge couplings in the simplest MSSM model, the best
(indirect) evidence we have so far for Supersymmetry. In considering
extensions of the MSSM, we will always require that this remarkable
feature is not spoiled. The addition of extra gauge singlets or complete
$SU(5)$ representations are the natural possibilities for extended
models where minimal unification is automatically preserved (at the
one-loop level). In our search for the upper Higgs mass limit we do not
restrict ourselves to these possibilities but consider other options.
Although splitted $SU(5)$ representations are difficult to arrange in GUT
models, they can easily arise in string models and may even help in
solving the mismatch between the MSSM unification scale and the String
scale [degenerate full $SU(5)$ multiplets do not modify (at one-loop)
the unification scale].

We assume that any model must contain at least the two minimal Higgs
doublets $H_{1,2}$ required to give quarks and leptons their masses. In
principle, more than two Higgs doublets could be involved in
$SU(2)_L\times U(1)_Y$ breaking, but in such a case a rotation in field
space can be made so that only $H_{1,2}$ have non-zero vacuum
expectation values.
Additional higher Higgs representations are generally very constrained
by $\rho=M_W^2/M_Z^2\cos^2\theta_W\simeq 1$. In addition, if they
contribute significantly to the
$W^\pm$ and $Z^0$ masses, the Higgs bound, sensitive to the doublet
contribution, gets weaker and, in addition, other light Higgses appear in
the spectrum \cite{coes}. It is thus conservative to assume that $\langle
H_{1,2}^0
\rangle$ are responsible for all the breaking and thus, the known $Z^0$
mass fixes $\langle H_{1}^0\rangle^2+\langle
H_{2}^0\rangle^2$$=v^2$.

To maximize the upper bound on $m_h$ we next assume that the model also
contains extra chiral multiplets with the appropriate quantum numbers to
give couplings of the form $W=h_X XH_iH_j$. Thus, $X$ can only be a
singlet ($S$) or a $Y=0,\pm1$ triplet ($T_Y$). From the gauge-invariant 
trilinear superpotential 
\bea
\label{superp}
W&=& \lambda_{1} H_1\cdot H_2 S
+  \lambda_{2} H_1\cdot T_0 H_2\nonumber\\
 &+&  \chi_{1} H_1\cdot T_1 H_1   +
 \chi_{2} H_2\cdot T_{-1} H_2, 
\eea
the tree-level mass bound follows (we correct here a normalization 
error in Refs.~\cite{eq92,eq93}):
\bea
\label{bound}
m_h^2/v^2&\leq&\frac{1}{2}(g^2+g'^2)\cos^2 2\beta +
(\lambda_{1}^2
+\frac{1}{2}\lambda_{2}^2)\sin^2 2\beta\nonumber\\
&+& 4\chi_1^2\cos^4\beta+ 4\chi_2^2\sin^4\beta.
\eea
The different dependence of the various terms with $\tan\beta$ will
make them important in different regimes. In particular we already 
anticipate that, in the large $\tan\beta$ region, the $\chi_2$ 
contribution will be crucial for the upper limit. $S$ and $T_0$ have the
same dependence while it can be shown that the effect of $\lambda_1$ is
always be more important than that of $\lambda_2$. For this reason we will
not take into account the possible effect of $T_0$ representations.

{\bf 2.} As the next step, one should impose triviality bounds on the
extra
couplings entering (\ref{bound}) by assuming they do not reach a Landau
pole below the unification scale. As stressed in \cite{gordy,alex}, large
values of the gauge couplings slow down the running of Yukawa couplings
with increasing energy and so they can be larger at the
electroweak scale. To get numerical values for these upper bounds we then
have to specify further the particle content of the model, imposing gauge
coupling unification and making the gauge couplings as large as possible.
For the renormalization group (RG) analysis we consider that all 
superpartners can be roughly characterized by a
common mass $M_{SUSY}$ below which the effective theory is just the SM
and, restricted by naturalness, we take $M_{SUSY}=1$ TeV. No such
constraint should be imposed on extra matter in vector like
representations, like ($5+\bar{5}$) $SU(5)$ pairs, which could be present
at intermediate scales. By setting their masses down to 1 TeV we are
enhancing their effect on the running of gauge couplings, which being
stronger will also tend to increase the $m_h$ bound. 

To achieve unification with only one scale $M_{SUSY}$ fixed to 1 TeV is
not completely trivial. When the MSSM is enlarged by one singlet $S$
and
a pair $\{T_1,T_{-1}\}$ (to cancel anomalies) the running $g_1^2=
5g'^2/3$ and $g^2$ meet at $M_X\sim 10^{17}$ GeV. Interestingly
enough, this is closer to the Heterotic String scale than the 
MSSM unification scale. Of course,
$g_3^2$ fails to unify unless extra matter is added. This can be
achieved, for example, by adding 4 ($3+\bar{3}$) [$SU(2)_L\times U(1)_Y$
`singlet quark' chiral multiplets] or one ($3+\bar{3}$) plus one
$SU(3)_c$ octet. In addition to this, one can still have one 
($5+\bar{5}$) $SU(5)$ pair, which will not change the unification scale. 
The unification of the couplings is shown in Figure~1 (solid lines).
\begin{figure}[hbt]
\centerline{
\psfig{figure=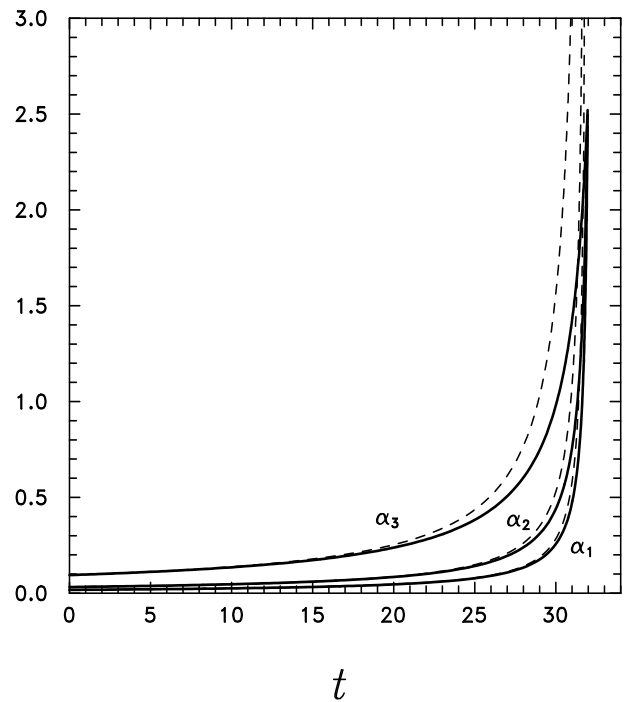,height=10cm,width=13.cm,bbllx=2.cm,bblly=3.cm,bburx=21.cm,bbury=17.cm}}
\caption{\footnotesize Running $\alpha_i$'s $(=g_i^2/4\pi)$ for the model
discussed in the text (solid lines) and upper perturbative limit (dashed
lines) with $t=\log(Q/M_{SUSY})$.}
\end{figure}

For comparison, dashed lines show the
running couplings when their beta functions are chosen in such a way that
all couplings reach a Landau pole at the unification scale. 
In this case the low-energy couplings are fully 
determined by the `light' matter content of
the model, which determines the RG beta functions.
This behaviour is dubbed non-perturbative
unification \cite{npu} and was long ago proposed as an alternative to 
conventional unification, with the attractive feature of having less
sensitivity of the low-energy couplings to high-energy physics (In
addition, the choice of $k_i$ normalization factors for gauge coupling
unification is now immaterial). The
dashed lines can be considered as the perturbative upper limit on the
gauge couplings and comparison with the solid lines shows that our model
is close to saturation and represents a concrete realization of the most
extreme scenario to maximize the $m_h$ bound. The emphasis here should lie
in this fact rather than in the plausibility or physics motivation of the
model per se. 

The particular model we use serves the purpose of illustrating
the fact that the Higgs mass bound can be saturated. 
The model includes exotic representations with non-canonical charge 
assignment, which can nevertheless appear in string models. $(3,1)_0$
and $(\bar{3},1)_0$'s can appear in general embeddings \cite{chl}
of the Standard Model group other than the usual embeddings in grand
unified groups like $SU(5)$ or $E_6$. On the other hand, $SU(2)$ triplets
are possible if the $SU(2)$ Ka\v c-Moody level is larger
than 1 (as we have seen, the effect of this on the unification condition 
is not important in the limiting case of non-perturbative unification).
While these triplets are the key ingredient to go beyond the Higgs 
mass limits of the MSSM, the additional representations included to
ensure unification are not uniquely determined. 
Different representations of exotic matter at intermediate scales, as
e.g color octets with canonical charge assignment, could
equally well give correct unification. In such cases, the lightest Higgs
can well be much heavier than in the MSSM even if the general upper
limit is not reached. 

Having optimized in this way the most appropriate running gauge
couplings, we
turn to the running of $\lambda_1$ and $\chi_{1,2}$. The relevant RG
equations can be found in Refs.~\cite{eq92,eq93}.
Inspection of this set of
coupled equations teaches that the low-energy values of each Yukawa
are maximized when other Yukawas (and self-interactions among $T_Y$'s and
$S$) are shut off. So, we consider each coupling in turn (setting the
others to zero, which is a RG-invariant condition) and compute its
maximum value at 1  TeV as a function of $\tan\beta$ (which influences
the top and bottom Yukawa couplings entering the RGs) for $M_t=175$ GeV.
This physical top-quark mass is related to the Yukawa coupling $h_t$ by
the $\overline{\mathrm MS}$ relation $h_t v
\sin\beta/\sqrt{2}=M_t/[1+4\alpha_s(Mt)/3\pi]$. Plugging the maximum
values of $\lambda_1$, $\chi_{1}$ or $\chi_2$ in the tree-level bound
(\ref{bound})
we obtain then three different upper limits on the tree-level Higgs mass.
To add the important radiative corrections we follow
the RG method, as explained e.g. in Refs.~\cite{rgmh1,rgmh2,rgmh3}, which
includes two-loop RG improvement and stop-mixing effects.

{\bf 3.} Before presenting our results, it is worth discussing in more
detail the bound presented in Eq.~(\ref{bound}). If the extra fields
responsible
for the enhancement of $m_h$ sit at 1 TeV, should their effect not
decouple from the low-energy effective theory? Indeed, in a simple toy
model
with an extra singlet $S$ coupled to $H_1\cdot H_2$ as in (\ref{superp}),
when a large supersymmetric mass is given to
$S$, the $F$-term contribution ($\sim \lambda_1^2$) to the Higgs doublet
self-interactions is cancelled by a tree diagram that interchanges the
heavy
singlet, thus realizing decoupling. If, on the other hand, we lower the
mass scale of the extra fields $S$ and $T_Y$ to the electroweak scale to
avoid decoupling, 
it is generically the case that more than one light Higgs appear in the
spectrum. A complicated mixed squared-mass matrix results whose lightest
eigenvalue
does not saturate the bound (\ref{bound}). Is then this mass limit simply
a too conservative overestimate of the real upper limit?
It is easy to convince oneself that, in the presence of soft breaking
masses, the perfect decoupling cancellation obtained in the large SUSY
mass limit does not take place (we are assuming here that SUSY masses,
if present for the extra matter, are not larger than 1 TeV) and the final
lightest Higgs mass depends in a complicated way on these soft
mass-parameters. The interesting outcome is that soft-masses can be
adjusted in order to saturate the bound (\ref{bound}) and so, the numbers
we will present can be reached in particular models and no limits
lower than these can be given without additional assumptions (which 
we will not make here, in the interest of generality).

\begin{figure}[hbt]
\centerline{
\psfig{figure=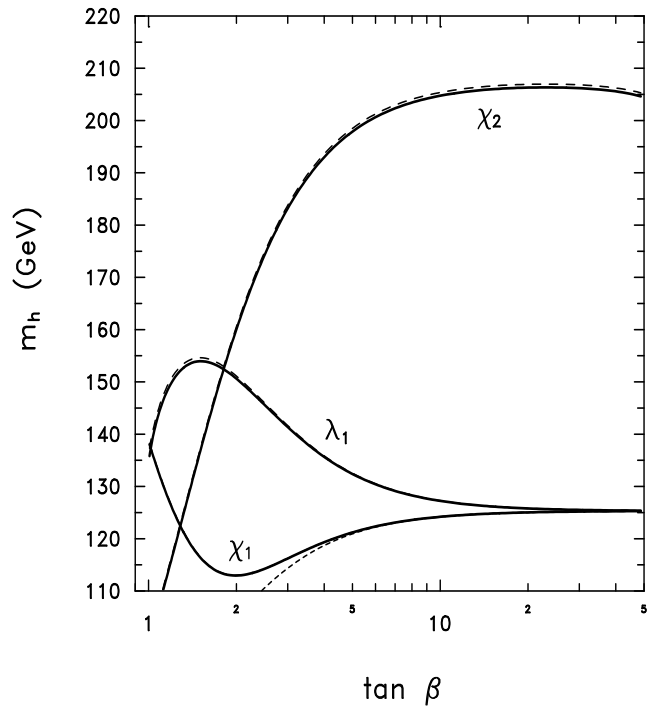,height=10cm,width=12.cm,bbllx=1.2cm,bblly=3.cm,bburx=20.2cm,bbury=17.cm}}
\caption{\footnotesize 
Radiatively corrected upper bounds on $m_h$ when different Yukawa
couplings are present in the 
model and for different assumptions on the running of gauge
couplings. The short-dashed line gives the upper bound in the MSSM.}
\end{figure} 
The final bounds, with radiative corrections included, are presented in
Figure~2. Solid lines are the mass limits when the particular model
described previously is assumed to determine the running of the gauge
couplings. Long-dashed lines are instead obtained when all gauge couplings
reach a Landau pole at $M_X\sim 10^{17} GeV$ and show that the particular
model we used practically saturates the absolute bounds. 

Below $\tan\beta\sim 1$ (beyond $\tan\beta\sim 60$), $h_t$ ($h_b$) 
reaches a Landau pole below $M_X$ and that
region is thus excluded. Lines labeled `$\lambda_1$' 
show the effect of the
singlet coupling $\lambda_1$. The maximal value of this coupling depends
on $\tan\beta$ through $h_t$. For small $\tan\beta$, $h_t$ is large and
forces $\lambda_1$ to be small. With increasing $\tan\beta$, $h_t$
becomes smaller and $\lambda_1$ can get larger values. This effect,
combined with the $\sin^22\beta$ dependence of the
$\lambda_1$-contribution to $m_h^2$ explains the shape of these lines. At
large $\tan\beta$, the $\lambda_1$ contribution to the mass limit shuts
off and the MSSM limit is recovered. [A limit similar to the $\lambda_1$-bound
has been recently obtained~\cite{alex} in the MSSM with a singlet field $S$
and pairs of $SU(5)$ $(5+\bar{5})$ saturating the gauge couplings.]
The same happens for the lines
labeled `$\chi_1$' which give the $\chi_1$-bound. The upper triviality bound
on this Yukawa coupling is basically independent of $\tan\beta$. The
interplay between the minimal ($\cos^2 2\beta$) piece of the bound 
(\ref{bound}) and the $\chi_1$ piece explains the shape of these lines.
It is interesting how the effect of the $\chi_2$ coupling is instead more
important for large $\tan\beta$, where it reinforces the minimal
contribution, providing the absolute upper limit. This is given by the lines
labeled `$\chi_2$' and can be as large as 205 GeV. We remark that these lines
assume that only one of the couplings $\lambda_1$, $\chi_1$ or $\chi_2$
is non-zero. If two couplings differ from zero simultaneously the bound
is reduced.

{\bf 4.} In conclusion, we calculate a numerical absolute upper limit on
the mass of the lightest supersymmetric Higgs boson for any model 
with arbitrary matter content compatible with gauge coupling unification 
around (and perturbativity up to) the
String scale. With this assumption, we show that this light Higgs mass 
can be as high as $\sim$ 200 GeV, significantly heavier than previously
thought. The model saturating this bound has asymptotically divergent
gauge couplings and points toward non-perturbative unification. 
Besides being of obvious interest to the experimentalists, this
result has interest for theorists too. If Higgs searches reach the MSSM
bounds without finding a signal for a Higgs boson, this could be taken,
if one is willing to stick to low-energy supersymmetry, as evidence for
additional matter beyond the minimal model. Without stressing the point
too much, this could be welcome to reconcile the unification scale with
the String scale.

\end{document}